\def\CP {\ensuremath{C\!P}\xspace}

\def\CP_T {\ensuremath{C\!P/T}\xspace}

\def\g {\ensuremath{\gamma}\xspace}

\def\BToXll {\ensuremath{B\rightarrow X_{s}\, \ell^{+}\ell^{-}}\xspace}
\def\epem {\ensuremath{e^+e^-}\xspace}
\def\BToXee {\ensuremath{B\rightarrow X_{s}\, \epem}\xspace}
\def\mumu {\ensuremath{\mu^+\mu^-}\xspace}
\def\BToXmm {\ensuremath{B\rightarrow X_{s}\, \mumu}\xspace}
\def\mll {\ensuremath{m_{\ell^+\ell^-}}}

\def\mx {\ensuremath{m_{X_s}}\xspace}

\def\epem {\ensuremath{e^+e^-}\xspace}

\newcommand{\lambdase}{\ensuremath{\tilde{\Lambda}_{78}}\xspace}

\def\Rhoz  {\ensuremath{\rho^0(770)} \xspace}
\def\mKpipi  {\ensuremath{m_{K  \pi  \pi}}\xspace}
\def\mKpi  {\ensuremath{m_{K \pi}}\xspace}

\def \CPBdecays  {\ensuremath{B \to f_{C\!P} \gamma}\xspace}
\def \BztoKstarRhozg  {\ensuremath{\Bz \to K^{*\pm}(\KS \pi^{\pm})\pi^{\mp} \g}\xspace}
\def \BztoKsRhozg  {\ensuremath{\Bz \to \KS\rho^0(\pi^{\mp} \pi^{\pm}) \g}\xspace}

\def\Seff {{\ensuremath{\mathcal{S}_{\KS\pip\pim\g}}}\xspace}
\def\Ceff {{\ensuremath{\mathcal{C}_{\KS\pip\pim\g}}}\xspace}

\def\Srho {{\ensuremath{\mathcal{S}_{\KS\rho\g}}}\xspace}
\def\Crho {{\ensuremath{\mathcal{C}_{\KS\rho\g}}}\xspace}
\def\D {{\ensuremath{\mathcal{D}_{\KS\rho\g}}}\xspace}

\def \splot {$_s\mathcal{P}lot$\xspace}

\def\babar{\mbox{\slshape B\kern-0.1em{\smaller A}\kern-0.1em
 B\kern-0.1em{\smaller A\kern-0.2em R}}\xspace}
\def\B{\ensuremath{B}\xspace}
\def\Bbar {\kern 0.18em\overline{\kern -0.18em B}{}\xspace}
\def\Bb {\ensuremath{\Bbar}\xspace}
\def\BB {\ensuremath{B\Bbar}\xspace} 
\def\Bz {\ensuremath{B^0}\xspace}
\def\Bzb {\ensuremath{\Bbar^0}\xspace}
\def\BzBzb {\ensuremath{\Bz {\kern -0.16em \Bzb}}\xspace}
\def\Bu {\ensuremath{B^+}\xspace}
\def\Bub {\ensuremath{B^-}\xspace}
\def\Bp {\ensuremath{\Bu}\xspace}
\def\Bm {\ensuremath{\Bub}\xspace}
\def\Bpm {\ensuremath{B^\pm}\xspace}

\def\BpBm {\ensuremath{\Bu {\kern -0.16em \Bub}}\xspace}

\def\piz {\ensuremath{\pi^0}\xspace}

\def\pip {\ensuremath{\pi^+}\xspace}
\def\pim {\ensuremath{\pi^-}\xspace}

\def\Kbar {\kern 0.2em\overline{\kern -0.2em K}{}\xspace}

\def\Kz {\ensuremath{K^0}\xspace}
\def\Kzb {\ensuremath{\Kbar^0}\xspace}
\def\KzKzb {\ensuremath{\Kz \kern -0.16em \Kzb}\xspace}
\def\Kp {\ensuremath{K^+}\xspace}
\def\Km {\ensuremath{K^-}\xspace}

\def\KpKm {\ensuremath{\Kp \kern -0.16em \Km}\xspace}
\def\KS {\ensuremath{K^0_{\scriptscriptstyle S}}\xspace}

\def\psitwos {\ensuremath{\psi{(2S)}}\xspace}
\def\Jpsi {\ensuremath{{J\mskip -3mu/\mskip -2mu\psi\mskip 2mu}}\xspace}
\mathchardef\Upsilon="7107
\def\Y#1S{\ensuremath{\Upsilon{(#1S)}}\xspace}

\newcommand{\tev}{\ensuremath{\mathrm{\,Te\kern -0.1em V}}\xspace}
\newcommand{\gev}{\ensuremath{\mathrm{\,Ge\kern -0.1em V}}\xspace}
\newcommand{\mev}{\ensuremath{\mathrm{\,Me\kern -0.1em V}}\xspace}
\newcommand{\kev}{\ensuremath{\mathrm{\,ke\kern -0.1em V}}\xspace}
\newcommand{\ev}{\ensuremath{\mathrm{\,e\kern -0.1em V}}\xspace}
\newcommand{\gevc}{\ensuremath{{\mathrm{\,Ge\kern -0.1em V\!/}c}}\xspace}
\newcommand{\mevc}{\ensuremath{{\mathrm{\,Me\kern -0.1em V\!/}c}}\xspace}
\newcommand{\gevcc}{\ensuremath{{\mathrm{\,Ge\kern -0.1em V\!/}c^2}}\xspace}
\newcommand{\gevcccc}{\ensuremath{{\mathrm{\,Ge\kern -0.1em V^2\!/}c^4}}\xspace}
\newcommand{\mevcc}{\ensuremath{{\mathrm{\,Me\kern -0.1em V\!/}c^2}}\xspace}

\def\invfb {\ensuremath{\mbox{\,fb}^{-1}}\xspace}

\def\btosll {\ensuremath{{b \to s \ell^+\ell^-}}\xspace}
\def\btosg {\ensuremath{{b \to s \gamma}}\xspace}

\def\mes        {\ensuremath{m_{\rm ES}}\xspace}
\def\DeltaE     {\ensuremath{\Delta E}\xspace}

\def \Nchanel  {\ensuremath{\Bz \to \KS \rho^0 \g}\xspace}
\def \NchanelPiPi  {\ensuremath{\Bz \to \KS \pip \pim \g}\xspace}
\def \Cchanel  {\ensuremath{\Bp \to \Kp \pim \pip \g}\xspace}

\documentclass[a4paper]{jpconf}
\usepackage{graphicx}
\RequirePackage{xspace}
\usepackage{relsize}

\begin{document}

\title{Penguin and rare decays in \babar}
\author{Simon Akar, on behalf of the \babar Collaboration}
\address{LPNHE, IN2P3/CNRS, Universit\'e Pierre et Marie Curie-Paris6, Universit\'e Denis Diderot-Paris7, F-75252 Paris, France\footnote{Now at CPPM, Aix-Marseille Universit\'e, CNRS/IN2P3, Marseille, France}}
\ead{simon.akar@lpnhe.in2p3.fr}


\begin{abstract}
We present recent results from the \babar Collaboration on radiative decays. 
These include searches for new physics via measurements of several observables such as the time-dependent $C\!P$ asymmetry in $B^0 \to \KS \pi^- \pi^+ \gamma$ exclusive decays, as well as direct $C\!P$ asymmetries and branching fractions in $B \to X_s \gamma$ and $B \to X_s \ell^+ \ell^-$ inclusive decays. 
\end{abstract}

\section{Introduction}
Until the end of 2007, the \babar\ experiment recorded \epem collisions at the \Y4S resonance with $471\times10^6$ \BB pairs produced, corresponding to an integrated luminosity of $429 \invfb$. 
Seven years after the end of the data-taking period, \babar is still producing many physics results. 
In the LHC era, \babar is still competitive, especially for channels involving neutral particles such as \piz or \KS.
We present here a selection of recent results on radiative decays from the \babar experiment. 
In the Standard Model (SM), both the \btosg and \btosll transitions are quark-level flavor-changing neutral current (FCNC) processes. Since all FCNC processes are forbidden at tree level in the SM, the lowest order diagrams representing these transitions must involve loops.
In such processes, QCD corrections can typically be described using an effective Hamiltonian defined as ${\rm H_{eff}} \propto \sum_{i=1}^{10} C_i\mathcal{O}_i$, where the $C_i$ and $\mathcal{O}_i$ are, respectively, the short-distance Wilson coefficients and local long-distance operators.
Contributions from New Physics (NP) in $b \to s$ transitions may modify the SM values of the Wilson coefficients, and change the values, predicted by the SM, of observables such as branching fractions and $C\!P$ asymmetries.

\section{Measurements of direct $C\!P$ asymmetries in $B \to X_s \gamma$ decays using a sum of exclusive decays}

In this analysis~\cite{Lees:2014uoa}, the direct $C\!P$ asymmetry, $A_{C\!P}$ for the sum of exclusive final states is measured:
\begin{equation}
	\label{eq:acpdef}
	A_{C\!P} = \frac{\Gamma_{\Bzb/\Bm \to X_{s}\gamma}-\Gamma_{\Bz/\Bp \to X_{\bar{s}}\gamma}}{\Gamma_{\Bzb/\Bm \to X_{s}\gamma}+\Gamma_{\Bz/\Bp \to X_{\bar{s}}\gamma}}~.
\end{equation}
The SM prediction for the asymmetry was found in a recent study to be in the range $-0.6\%<A_{C\!P}^{SM}<2.8\%$~\cite{Benzke:2010tq}. From Ref.~\cite{Benzke:2010tq}, another test of the SM is proposed, via the measurement of the difference in $A_{C\!P}$ in charged and neutral $B$ mesons: 
\begin{equation}
\Delta A_{C\!P} = A_{\Bpm \to X_s\gamma} - A_{\Bz/\Bzb \to X_s\gamma} \simeq 0.12 \times \frac{\tilde{\Lambda}_{78}}{100\mev} \mathrm{Im}\left(\frac{C_{8g}}{C_{7\gamma}}\right)~,
\label{eq:dacpim78}
\end{equation}
where $C_{7\gamma}$ and $C_{8g}$ are the Wilson coefficients corresponding to the electromagnetic dipole and the chromo-magnetic dipole transitions, respectively, and $\tilde{\Lambda}_{78}$ is the interference amplitude.
Since in the SM, both $C_{7\gamma}$ and $C_{8g}$ are real, $\Delta A_{C\!P}$ is expected to be zero. 

The $B$ meson decays are fully reconstructed in 16 self-tagging final states, which are listed in Ref.~\cite{Lees:2014uoa}. 
The raw asymmetry is extracted from a simultaneous fit to the \textit{energy-substitued} \B meson mass, $m_{\rm ES}$, distributions of \B and \Bb tagged samples; $m_{\rm ES} \equiv \sqrt{(\sqrt{s}/2)^2 - (p^*_\B)^2} $, where $p^*_{\B}$ is the momentum vector of the \B in the \epem center-of-mass (CM) frame and $\sqrt{s}$ is the total energy of the \epem system. 
The direct $C\!P$ asymmetry is obtained after correcting the raw asymmetry for detector effects. 
Possible dilutions from the presence of peaking backgrounds in the \mes distributions are taken into account as systematic uncertainties. 
The measured value is $A_{C\!P} = +(1.73\pm1.93\pm1.02)\%$, where the first quoted error is statistical and the second is systematic. 
This corresponds to the most precise measurement to date and is compatible with the SM. 
The measurement of $\Delta A_{C\!P}$ is obtained by a simultaneous fit to the separate charge and neutral \B samples, such as $\Delta A_{C\!P} = +(5.0\pm3.9\pm1.5)\%$, where the first quoted error is statistical and the second is systematic.

\begin{figure}[h!]
\centering
\begin{tabular}{cc}
	\includegraphics[height= 4.5cm] {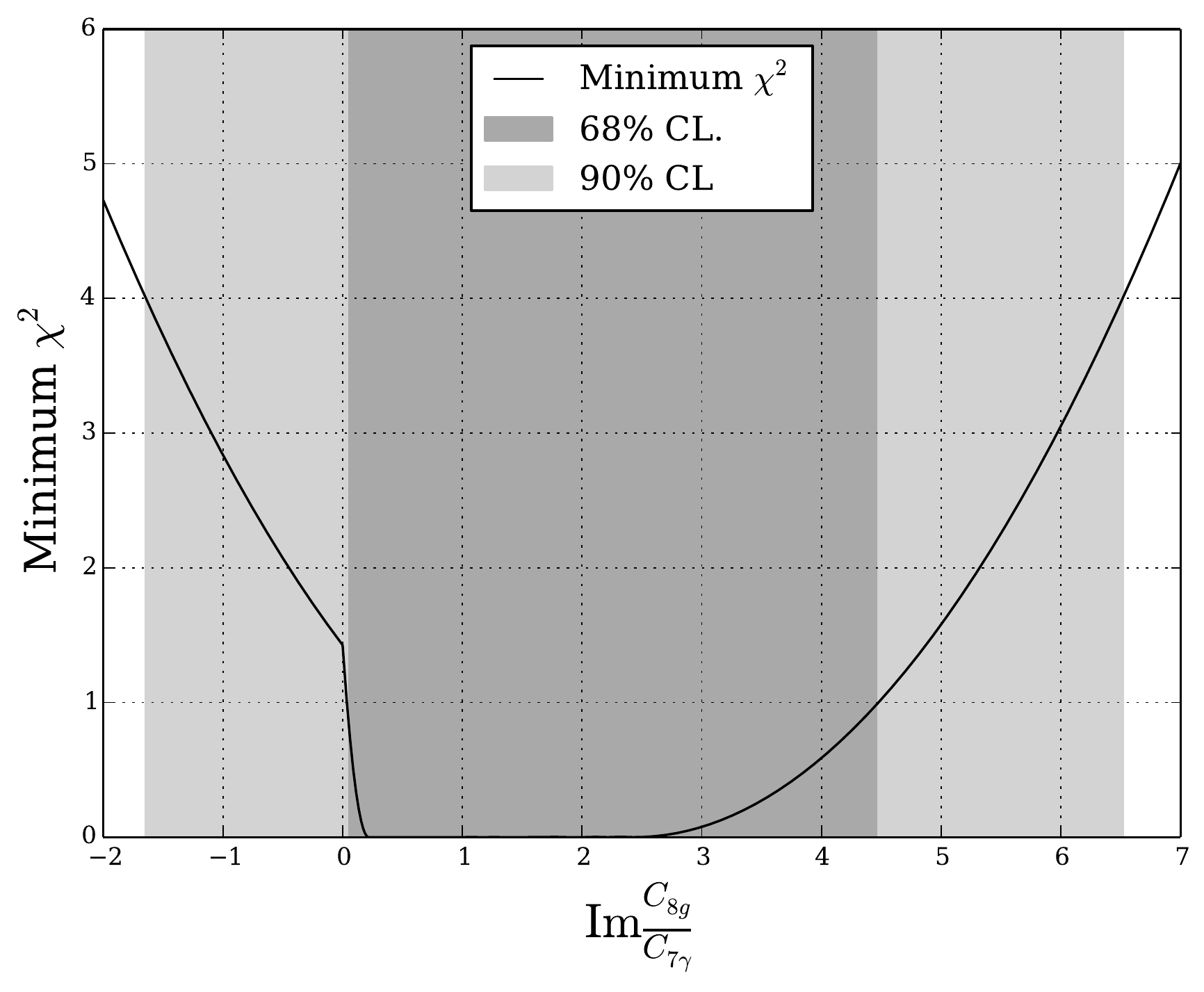}
	&
	\includegraphics[height= 4.5cm] {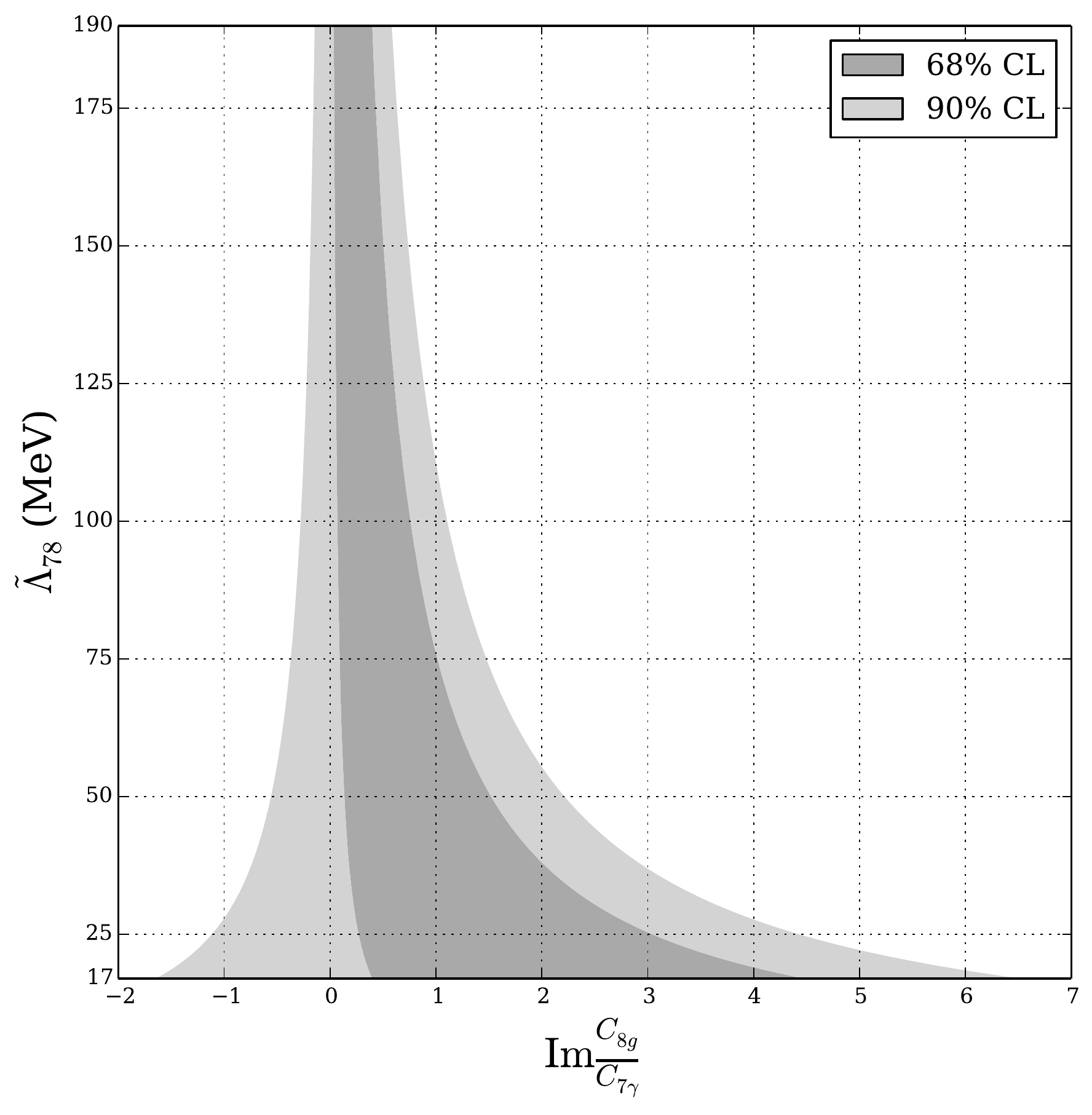}\\
\end{tabular}
\vspace{-10pt}
\caption{
 (left) The minimum $\chi^2$ for given $\mathrm{Im}(C_{8g}/C_{7\gamma})$ from all possible values of $\lambdase$. 68\% and 90\% confidence intervals are shown in dark gray and light gray, respectively.
(right) The 68\% and 90\% confidence intervals for  $\mathrm{Im}(C_{8g}/C_{7\gamma})$ as a function of $\lambdase$.
}
\label{fig:limits}
\vspace{-17pt}
\end{figure}

The interference amplitude, $\tilde{\Lambda}_{78}$, in Eq.~\ref{eq:dacpim78} is only known as a range of possible values: $17\mev < \tilde{\Lambda}_{78} < 190 \mev $. Using the measured values of $\Delta A_{C\!P}$, a $\chi^2$  minimization is performed for given $\mathrm{Im}(C_{8g}/C_{7\gamma})$ from all possibles values of $\tilde{\Lambda}_{78}$, as shown in Fig.~\ref{fig:limits}. The plateau of minimum $\chi^2 = 0$ corresponds to the region where a value of $\tilde{\Lambda}_{78}$ always exists in the allowed range such as the theoretical value of $\Delta A_{C\!P}$ matches exactly the experimental one. 
The 68\% and 90\% confidence limits are then obtained from the ranges of $\mathrm{Im}(C_{8g}/C_{7\gamma})$, which yield the minimum $\chi^2$ less than 1 and 4, respectively, such as: $0.07 \leq \mathrm{Im}( C_{8g}/C_{7\gamma}) \leq 4.48,\textrm{ at 68\% CL}$ and $-1.64 \leq \mathrm{Im}( C_{8g}/C_{7\gamma}) \leq 6.52, \textrm{ at 90\% CL}$.

\section{Measurement of the \BToXll branching fraction and search for direct $C\!P$ violation from a sum of exclusive final states}

In the present analysis, the inclusive decay \BToXll is studied, where $X_s$ is a hadronic
system containing exactly one kaon,
using a sum over 20 exclusive final states, where $\ell^+\ell^-$ is either $e^+e^-$ or $\mu^+\mu^-$, as listed in Ref.~\cite{Lees:2013nxa}.

The total branching fraction (BF), as well as partial
BFs in five disjoint dilepton mass-squared $q^{2} \equiv \mll^2$ bins and four hadronic mass $\mx$ bins, which are defined in Ref.~\cite{Lees:2013nxa}. Events with $q^{2}$ corresponding to signal-like charmonium backgrounds \Jpsi and \psitwos from $B$ decays are rejected.
After requiring the invariant mass of the hadronic system $\mx < 1.8 \gevcc$, the entire selection represents $\sim 70\%$ of the inclusive \BToXll rate. The missing hadronic final states, as well as states with $\mx > 1.8 \gevcc$ are accounted for
using JETSET fragmentation~\cite{Sjostrand:1993yb} and
theory predictions.

The signal yields are extracted, in each $q^{2}$ and $\mx$ bins, from a simultaneous fit to $m_{\rm ES}$ and a likelihood ratio $L_R$, which is defined as $L_R \equiv
\mathcal{P}_S / (\mathcal{P}_S + \mathcal{P}_B )$ with $\mathcal{P}_S$ $(\mathcal{P}_B)$ the probability for a correctly-reconstructed signal (\BB background) event calculated based on the response of boosted decision trees.

The total BF for $q^{2} > 0.1 \gevcccc$ is measured to be $\mathcal{B}(\BToXll) = 6.73^{+0.70}_{-0.64}{}^{+0.34}_{-0.25} \pm 0.50 \times 10^{-6}$, where the first uncertainties are statistical, the second experimental systematics and the third model-dependent systematics. The same convention for the quoted uncertainties is used hereafter. This result is less than $2\sigma$ above the SM prediction of $\mathcal{B}_{SM} = 4.6 \pm 0.8 \times 10^{-6}$~\cite{Ghinculov:2003qd}. 

In the low mass range, for $1 < q^{2} < 6 \gevcccc$, the BF is measured to be $\mathcal{B}^{\rm low}(\BToXll) = 1.60^{+0.41}_{-0.39}{}^{+0.17}_{-0.13} \pm 0.18 \times 10^{-6}$, which is in good agreement with the SM predictions of ${\cal B}^{\rm low}_{SM}(\BToXmm) = (1.59 \pm 0.11) \times 10^{-6}$ and ${\cal B}^{\rm low}_{SM}(\BToXee) = (1.64 \pm 0.11) \times 10^{-6}$~\cite{Huber:2007vv}. 
In the high mass range, for $q^{2} > 14.2 \gevcccc$, the BF is measured to be $\mathcal{B}^{\rm high}(\BToXll) = 0.57^{+0.16}_{-0.15}{}^{+0.03}_{-0.02} \pm 0.00 \times 10^{-6}$, which is about $2\sigma$ higher than the SM predictions for $q^{2} > 14.4 \gevcccc$ of
${\cal B}^{\rm high}(\BToXmm) = (0.24 \pm 0.07) \times 10^{-6}$ and
${\cal B}^{\rm high}(\BToXee) = (0.21 \pm 0.07) \times 10^{-6}$~\cite{Huber:2007vv}.
The measured partial BFs results in bins of $q^{2}$ and $\mx$ are detailed in Ref~\cite{Lees:2013nxa}.

A search for the direct $C\!P$ asymmetry in \BToXll decays, using the 14 self-tagging final states as listed in Ref.~\cite{Lees:2013nxa}, is also performed. For $q^{2} > 0.1 \gevcccc$, the measured $C\!P$ asymmetry, $A_{C\!P}(\BToXll)=0.04 \pm 0.11 \pm 0.01$, is consistent with the SM prediction, where it is expected to be suppressed well below the $1\%$.

%
%
%

\section{Time-dependent analysis of $B^0 \to \KS\pip\pim\gamma$ and studies of the $\Kp\pim\pip$ system in \Cchanel decays}

In $b \to s\g$ transitions, the SM predicts that \Bz(\Bzb) decays are related predominantly to the presence of right- (left-) handed photons in the final state. Therefore, the mixing-induced $C\!P$ asymmetry in \CPBdecays decays, where $f_{C\!P}$ is a $C\!P$ eigenstate, is expected to be small. 
%

One of the goals of the present study is to perform a time-dependent analysis of \NchanelPiPi decays to 
extract the direct and mixing-induced
$C\!P$ asymmetry parameters, \Crho and \Srho, in the \Nchanel mode.
However, due to the large natural width of the $\Rhoz$, a non negligible amount of \BztoKstarRhozg events, which do not contribute to \Srho, are expected to lie under the $\Rhoz$ resonance and dilute \Srho. We can define a dilution factor \D such as
$\D = {\Seff}/{\Srho}$,
where \Seff is the effective value of the mixing-induced $C\!P$ asymmetry measured for the whole \NchanelPiPi dataset. 
Since a small number of signal events is expected in this sample, it is difficult to discriminate \BztoKstarRhozg from \BztoKsRhozg decays. Hence the dilution factor needs to be obtained by other means. 
To do that, the amplitudes of the different resonant modes are extracted in the charged decay channel\footnote{Charge conjugation is implicit throughout the document.} \Cchanel, which has more signal events and is related to \NchanelPiPi by isospin symmetry. 
Assuming that the resonant amplitudes are the same in both modes, the dilution factor is calculated from those of \Cchanel.
Moreover, the branching fractions of the different $\B \to K_{\footnotesize{\textrm{res}}} \g$ intermediate decay modes (where $K_{\footnotesize{\textrm{res}}}$ designates a kaonic resonance decaying to $K\pi\pi$) are in general not well known. 
We also use the amplitude analysis of the charged decay mode \Cchanel to extract them.

In the charged \B-meson decay mode, we first perform a fit to data to extract the yield of \Cchanel signal events.
The fit uses the knowledge of three discriminating variables, \mes, \DeltaE and a Fisher discriminant output, to discriminate signal events from backgrounds.  
\DeltaE is defined as the difference between the expected and reconstructed \B meson energy, $\Delta E \equiv E^*_{\B}-\sqrt{s}/2$, where $E^*_{\B}$ is the reconstructed energy of the \B in the \epem CM frame.
Using information from the maximum likelihood fit, the $K\pi\pi$, $K\pi$ and $\pi\pi$ invariant mass spectra in signal events are extracted using the \splot technique~\cite{sPlot}. 
Then fitting the \mKpipi and \mKpi spectra, the amplitudes and BFs of the kaonic resonances and the intermediate state resonances, respectively, are extracted, allowing to compute the dilution factor, such as $\D = 0.549^{+ 0.096}_{-0.094}$, where the quoted uncertainties are sums in quadrature of statistical and systematic uncertainties.

\begin{figure}[t!]
\vspace{-13pt}
\centering
\begin{tabular}{cc}
	\includegraphics[height= 4.0cm] {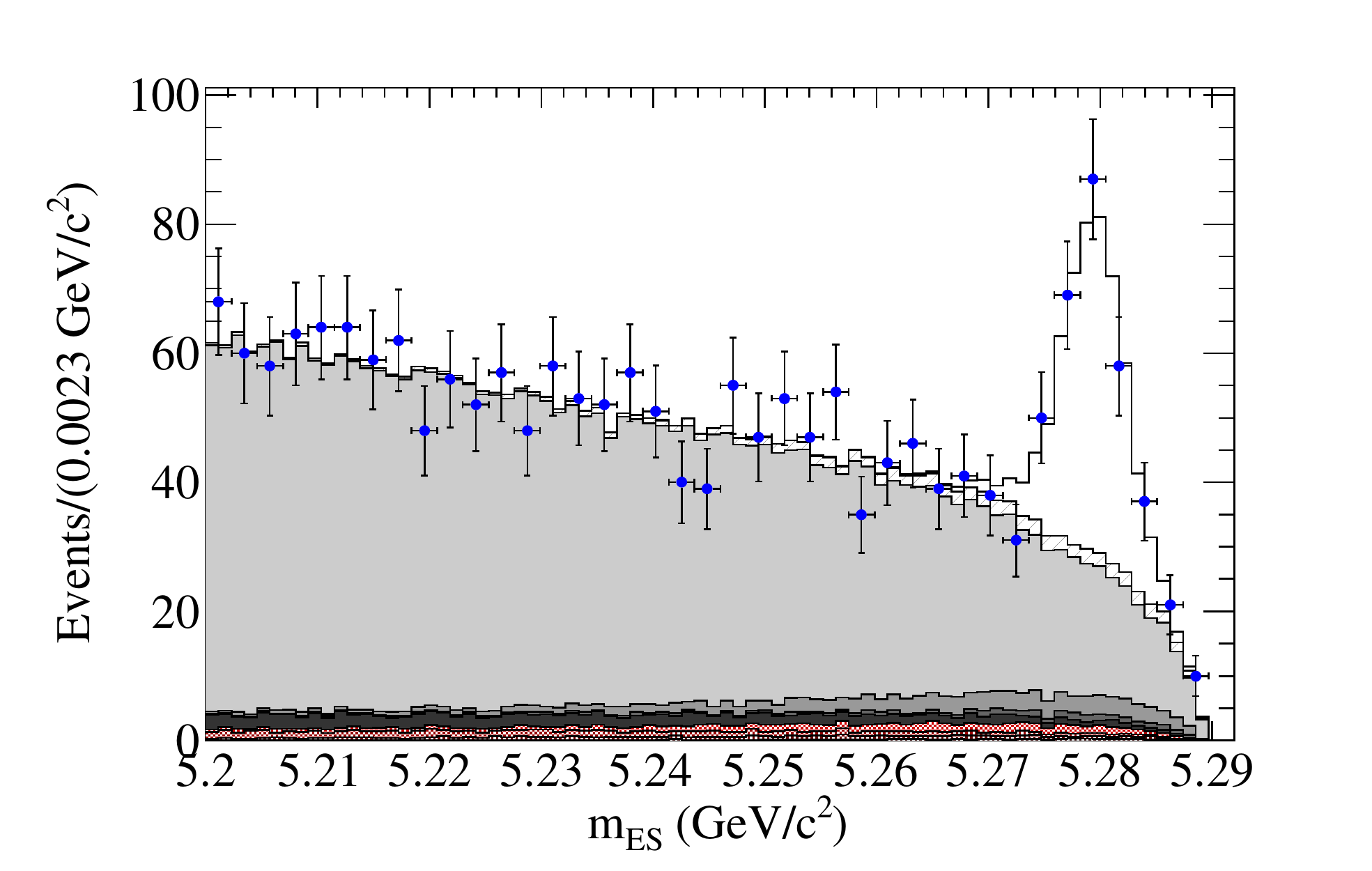}
	&
	\includegraphics[height= 4.0cm] {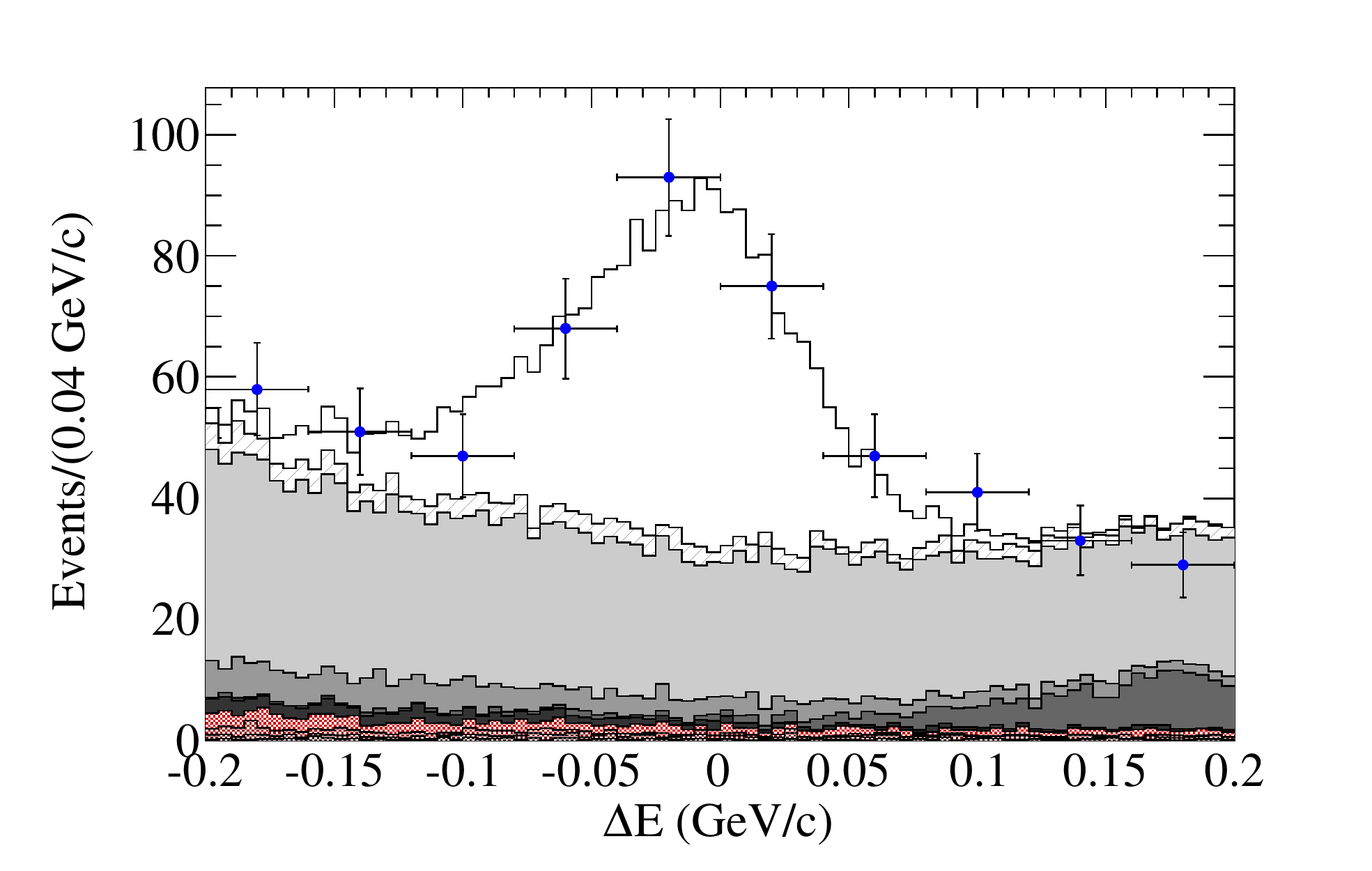}\\
\includegraphics[height= 4.0cm] {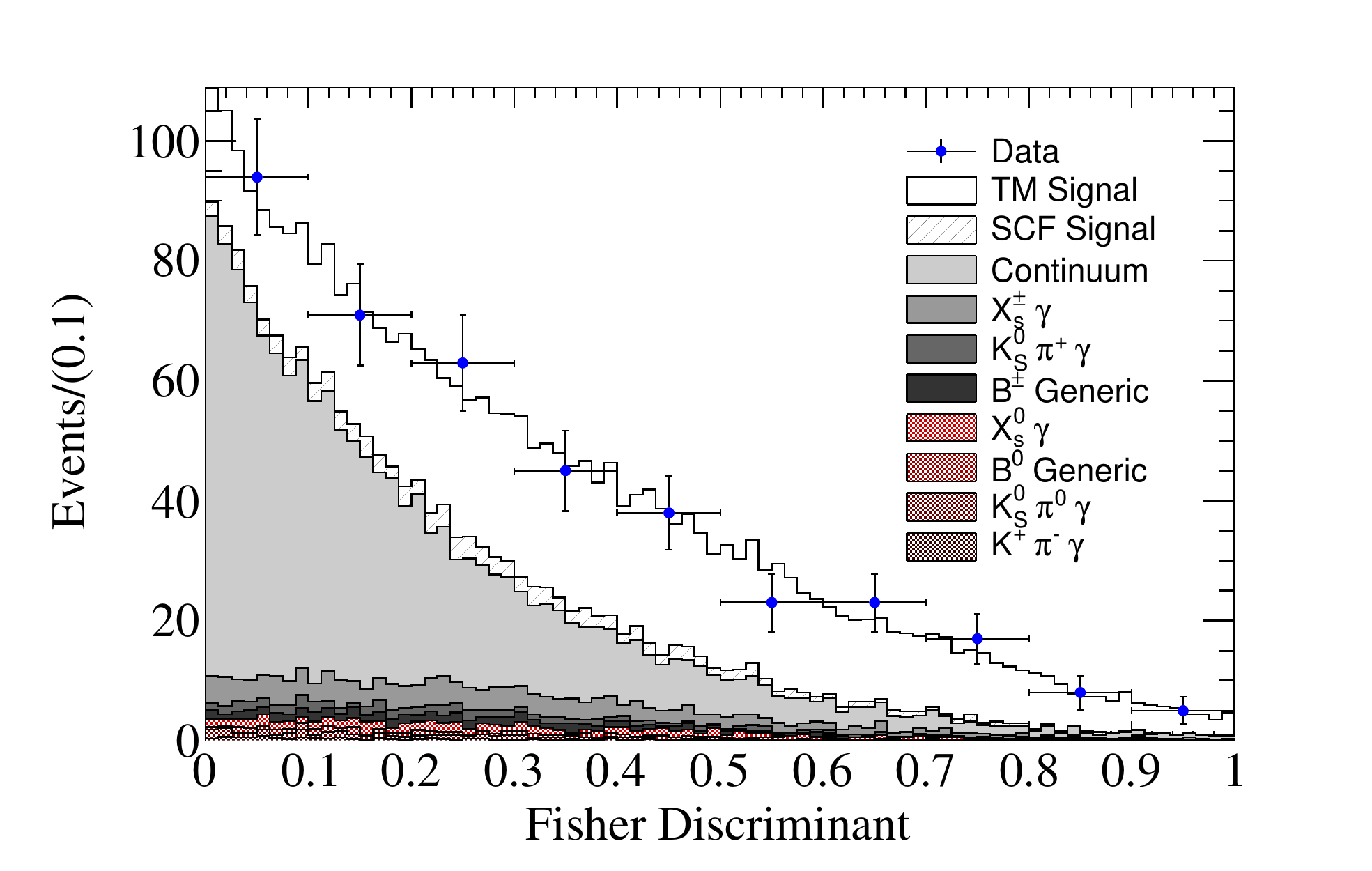}	
&
	\includegraphics[height= 4.0cm] {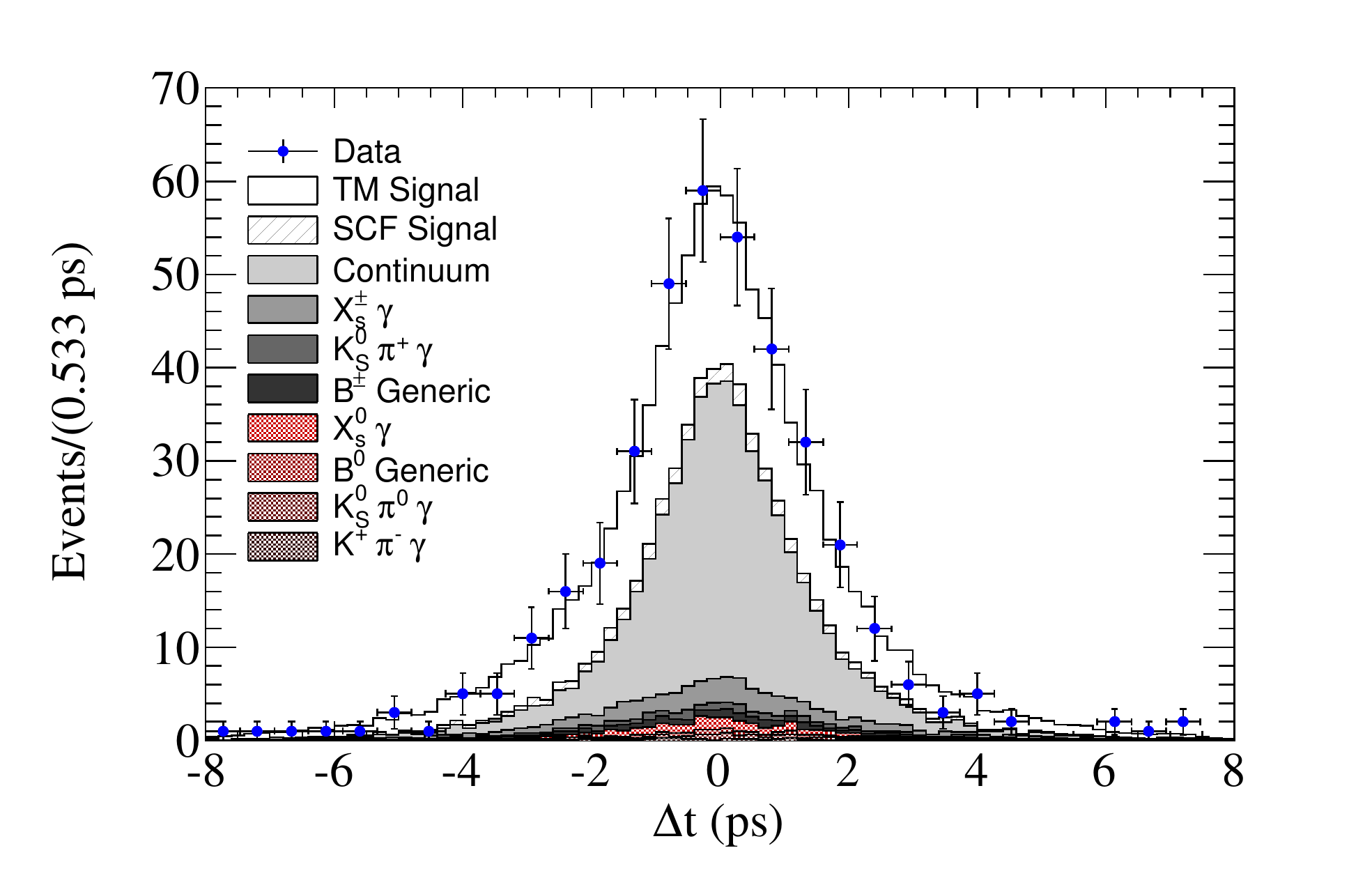}\\
\end{tabular}
\vspace{-8pt}
\caption{
Distributions of \mes (top left), \DeltaE (top right), the Fisher-discriminant output (bottom left), and $\Delta t$ (bottom right), showing the fit results on the \NchanelPiPi data sample.
The distributions have their signal/background ratio enhanced by means of the following requirements:
$ -0.15 \leq \DeltaE \leq 0.10 \gevc$ (\mes), $\mes > 5.27 \gevcc$ (\DeltaE) and $\mes > 5.27 \gevcc \,;\, -0.15 \leq \DeltaE \leq 0.10 \gevc $ (Fisher and $\Delta t$).
}
\vspace{-13pt}
\label{fig:FitProjNeutral}
\end{figure}

In the neutral \B-meson decay mode, we perform a fit to data to extract the effective $C\!P$ asymmetry parameters, using four discriminating variables, \mes, \DeltaE and a Fisher discriminant output and the proper-time difference $\Delta t$. Fig.~\ref{fig:FitProjNeutral} shows the fit projections.
We obtain the $C\!P$-violating parameters $\Seff = 0.137 \pm 0.249^{+ 0.042}_{-0.033}$ and $\Ceff= -0.390 \pm 0.204^{+ 0.045}_{-0.050}$, where the first quoted errors are statistical and the second are systematic. Using the dilution factor, extracted from the charged mode analysis, we extract the time-dependent $C\!P$ asymmetry related to the hadronic $C\!P$ eigenstate $\rho^0 \KS$ and obtain $ \Srho = 0.249 \pm 0.455^{+ 0.076}_{-0.060}$, which is compatible with the SM expectation of $\sim 0.03$~\cite{Atwood:2004jj}.

\section{Conclusions}
The \babar Collaboration, seven years after the shutdown of the experiment, is still producing competitive results.
Three recent analyses have been presented, for which all the results are in agreement with the SM predictions. 


\end{document}